\documentstyle[12pt,epsfig]{article}
\begin{document}
\baselineskip=2pc
\def\bfg #1{{\mbox{\boldmath $#1$}}}
\begin{center}
{\large \bf { ROLE OF DEUTERON ${\bf NN^*}$-COMPONENTS
IN PROCESSES  ${\bf pd \to dp}$ AND
  ${\bf pd\to dN^*}$} }

 Yu.N. Uzikov\footnote{e-mail address: uzikov@nusun.jinr.dubna.su\\
telephone: +7 09621 62417\\
FAX: +7 09621 66666}

{\it Laboratory of Nuclear Problems,
Joint Institute for Nuclear Research,\\
 Dubna, Moscow reg., 141980 Russia}
\end{center}

\vskip 1em
\section*{Abstract}

The contribution of nucleon isobar
  $N^*$ exchanges to backward elastic pd-scattering is calculated
on the basis of deuteron 6q-model and found to be negligible in comparison
with the neutron exchange.
It is shown that the pole amplitude of neutron pickup from the
deuteron $nN^*$-component is favoured in the reaction $pd\to dN^*$
for backward going $N^*(1440)$ and $N^*(1710)$ at kinetic energy of
 incident proton of 1.5--2 GeV  whereas the  triangular diagram with subprocess $pp\to
d\pi^+$ related to the usual $pn-$component
of deuteron is  considerable suppressed.

\vspace {1cm}
PACS 25.10.+s; 25.40.Cm; 24.85.+p

keywords: nucleon isobars in deuteron; $pd$-interaction
\newpage

\section{Introduction}

 An idea of preexistence of nucleon isobars  in the deuteron at
 short  NN-distances suggested for the first time in Ref. \cite{KK} is
 compatible   both  with  the meson exchange theory \cite{aren}
 and 6-quark picture of the deuteron structure \cite {stch}. Backward elastic
 pd-scattering, $pd\to dp$, is one source of information on the short-range
 structure of the  deuteron. According to calculations \cite {KK} based
 on the Regge phenomenology
and analysis \cite {3} performed in the meson exchange theory
 the contribution of the $NN^*$-component to the $pd\to dp$ process is essential
to explain the experimental data at energies $\sim 1 GeV$.
 However, the application of the Regge-model at rather low
energies  as well as  considerable uncertainties in  knowledge on
 $meson-N-
 N^*$ vertices  make these estimations questionable. Developed in last
 decade,  the 6-quark model of the deuteron \cite {2}-\cite{glozkuch}
 provides  a new regular
 approach  to  construction of $dNN^*$- vertices. In this model the deuteron
 structure at
 short relative NN-distances $r_{NN}\leq 1 fm$ is determined by superposition
 of nonexcitated $s^6$ and excitated $s^4p^2-s^52s$  6-quark shell-model
 configurations.
  Presence of
 two-quantum excitations in the configuration $s^4p^2-s^52s$ is a reason for
  the  phenomenological repulsive core in the NN-interaction potential
  \cite {core}. Besides, the excitated quark configuration  leads
 to an admixture of a small $NN^*$-component in the deuteron wave function.
The  effective numbers and momentum distributions are calculated in the
 framework  of this approach  \cite{glozkuch, ksg}. Recently the results
 \cite{ksg} for $dNN^*$ vertices were found sufficient \cite {kob96} to
explain the available experimental data  on the inclusive reaction
of deuteron disintegration $d+A\to p(0^o)+X$\cite{ableev} within the
$n+N^*$-exchange mechanism.

   In this work the contribution of $N^*$- exchanges to $pd\to dp$
  (Fig.1,a) is calculated in the interval of incident proton kinetic energy in
  the labsystem of $T_p=0.5-3GeV$ on the basis of the 6-quark model
  \cite{glozkuch,ksg}
  for $dNN^*-$ vertices. As is found here, this contribution
  is negligible  in comparison with the mechanism of neutron exchange
calculated in the Born approximation
(Fig.1, b)
 and with account of rescatterings (Fig.1,c-e).
  Furthermore we investigate the reaction $pd\to dN^*$ for the backward going
   $N^*$-isobar in the framework of the neutron exchange (NE) pole diagram
   (Fig.1,f) and
  triangle diagram (Fig.1,g) of one-pion exchange (OPE).
  The experimental investigation of the $pd\to dN^*(1440)$ reaction is
planned at SATURNE \cite{16}.
   If the NE-mechanism
  dominates, this reaction can give the direct information on the deuteron
  $nN^*$-component. The OPE amplitude involves the usual $np$-component of
  the deuteron and masks the $NN^*$-component. However, as  will be shown
 here, for the nucleon-like $N^*(1/2^+)$-states
  there is a kinematic region for the $pd\to dN^*$ reaction in which
  the OPE   mechanism is considerably suppressed.

\section {The model}

   The relativistic effects play an important role in the NE-mechanism at
   energies $\geq 1GeV$, especially for the $d\to p+N^*$ channel with
  large binding
   energy, $\varepsilon\sim 500 MeV$ \cite{uz92}. In order to allow for
relativistic effects we use here the phenomenological relativistic approach
for the three-body problem developed in Ref. \cite{bkt}. In this
the amplitude of the process $pd\to dB$, where $B$ denotes either
a  proton (for $pd\to dp$) or $N^*$ (for $pd\to dN^*$), in the framefork of one
baryon exchange (OBE) can be written as direct generalization of the
${pd\to dp}$ formalism of Ref. \cite {klsh}
\begin{equation}
\label{gl1}
 A_{OBE}=4 \sqrt{E_{d}(E_{p}+E_N)E_{d'}(E_{B}+E_N)}\,
{ \sqrt{s}-M_0 \over {E_N}}
\sum _{|N>} \left \{ \Psi_{\lambda '}^{\sigma _p \sigma _N}
({\bf q}')\right \}^+
\Psi_{\lambda }^{\sigma_B\sigma_N}({\bf q}).
\end{equation}
Here $E_k=\sqrt{m_k^2+{\bf p}^2_k}$ and ${\bf p}_k$
 are the energy and momentum of the k-th  particle in the p+d c.m.s.,
 $m_k$ is its mass; $M_0=E_N+E_{p}+E_{B}$; $\sqrt{s}$ is the invariant
mass of   the
 $p+d=d+B$ system;
 $\Psi_{\lambda '}^{\sigma_p \sigma_N}
 (\Psi_{\lambda }^{\sigma_B \sigma_N}$) is the deuteron wave function in the
 channel $d\to Np\, (d\to NB)$ normalized to the effective number,
 $N_{pN}^d $, of the corresponding channel
\begin{equation}
{1\over 2J_d+1} \sum _{\lambda , \sigma _p, \sigma _N}
\int |\Psi_{\lambda }^{\sigma _p, \sigma _N}({\bf q})|^2\;
{\rho_{pN}^{-1}(q) }\, {d^3q\over {(2\pi )^3}}= N_{pN}^d,
\end{equation}
where $\rho_{pN}(q)= 2{\varepsilon_p(q) \varepsilon_N(q)}/
[\varepsilon_p(q)+\varepsilon_N(q)]$; $\varepsilon_k({\bf q})=
\sqrt{m_k^2+{\bf q}^2_k}$; $\sigma _j$ is the spin projection
of the nucleon $j$ (j=p,B,N); $\lambda \, (\lambda ')$ denotes the spin
 projection of the  initial (final) deuteron.  The sum over the internal states
$\varphi_N$, including $\sigma_N$, of the transferred baryon  N
 (neuteron or $N^*$) is assumed in  Eq. (\ref{gl1}).
 The combinator factor $(\sqrt{2})^2$ is included in
Eq.({\ref {gl1}) since the 6-quark deuteron wave function is fully
 antisymmetric.   The
 arguments ${\bf q}$ and ${\bf q}'$ of the initial and final deuteron wave
 functions  can be written in the following form
\begin{equation}
\label{gl3a}
{\bf q}'={\bf p}_p-{\varepsilon_{p}({\bf q}')+E_{p}\,
\over\varepsilon_N({\bf q}')+E_N+\varepsilon_{p}({\bf q}')+E_{p}}\,{\bf d}',
\end{equation}
\begin {equation}
\label {gl3b}
{\bf q}={\bf p}_B-{\varepsilon_{B}({\bf q})+E_{B}\,
\over\varepsilon_N({\bf q})+E_N+\varepsilon_{B}({\bf q})+E_{B}}\,{\bf d},
\end{equation}
 the relations
${\bf p}_N={\bf d}-{\bf p}_{B}={\bf d}'-{\bf p}_{p}$
are used here which are valid in the p+d c.m.s. \cite{bkt};
${\bf d} \,({\bf d}')$ is the momentum of the initial (final) deuteron.
The amplitude (\ref{gl1}) is related to the c.m.s. cross section of
$pd\to dB$ as
\begin{equation}
\label{lgsec}
{d\sigma\over  d\Omega }= {1\over 64\pi^2 s} {p_B\over p_p}{\overline
{|A|^2}}.
\end{equation}

The basis for calculation of $dNN^*$-vertices is the fully antisymmetric
 6q-wave function of
deuteron which in the resonating group method (RGM) has a form
\begin{equation}
\label{gl24}
\Psi_{d}(1,\dots,6)={\hat A}\{\varphi_p(1,2,3)\varphi_n(4,5,6)\chi ({\bf r})\}.
\end{equation}
Here $\varphi_p$ and $\varphi_n$ are the quark wave functions of proton and
neuteron, $\chi({\bf r})$ is the RGM distribution function for the $pn$
 component of deuteron and ${\hat A}$ is the quark antisymmetrizer.
 When deriving the function $\chi({\bf r})$ one can either calculate it
in the microscopic 6q-dynamics or construct it by means of  RGM-renormalization
 procedure \cite{glozkuch} for the conventional
phenomenological wave function of deuteron in pn-channel, like Paris
or RSC. The difference between the effective numbers for these two methods
is negligible, of few percentage \cite{glozkuch}.
The translationally invariant shell model (TISM) state
 is  used as the iternal state of  quark motion in the baryon.
 The wave function
 $\Psi_\lambda^{\sigma_B \,\sigma_N}$
  for the channel
$d\to N+B$ entering  Eq.(\ref{gl1}) is determined by the overlap integral
 between the 6-quark wave function of the deuteron, $\Psi_{d}$, (\ref{gl24})
   and the product of the internal wave functions of the baryons,
  $\varphi_N$ and $\varphi_{B}$,
 as  $\Psi_\lambda^{\sigma_B \,\sigma_N}= \sqrt{6!\over 3!3!2}
  <\varphi_N\varphi_B|\Psi_{d}>$.
The details of the formalism and the effective numbers for $N^*$ in the
 deuteron are  presented in Refs. \cite{glozkuch,ksg}.

 Rescatterings in the initial and final states for the
  NE amplitude are taken into account here in the eikonal approximation on
 the  basis of the method
 developed in Ref. \cite{blu}.
 As a result, besides  the Born term (Fig.1,
 a or b), three additional terms arise allowing for pd-rescattering at small
 angles in the initial state (Fig.1,c), pp-rescattering in the final state
 (Fig.1, d) and rescatterings both in the initial and final states simultaneously
 (Fig.1,e).

 The spin-averaged square of the NE-amplitude of the $pd\to dN^*$ reaction
 (Fig.1,f) takes the form
\begin{eqnarray}
\label{pl5}
{\overline {|A_{NE}(pd\to dN^*)|^2}}={3\over 64\pi^2}\,K^2\rho_{pn}(q')\,
\rho_{nB}(q)
\,[u^2(q')+w^2(q')]\
\Phi^2_{N_BL_B}(q),
\end{eqnarray}
where
$K$ is the same kinematic factor as in front of the sum sign in Eq.(\ref{gl1}),
$u$ and $w$ are the S- and D-components of the deuteron function in the
$d\to pn$ channel, $\Phi^2_{N_BL_B}(q)$ is the momentum distribution in the
channel $d\to nN^*$ for the $N^*$-isobar with the
number of internal excitation quanta $N_B$
and  internal  orbital momentum  $L_B$
normalized by the condition
$\int_o^\infty\Phi^2_{N_BL_B}(q)q^2\, dq=N_{nB}^d(2\pi)^3$.
The corresponding formula for the NE mechanism in $pd\to dp$ follows from
Eq.(\ref{pl5}) after substitution $B\to p$,
 $\Phi^2_{N_BL_B}(q)\to u^2(q)+w^2(q)$.
In the framework of the NE-mechanism the tensor polarization of the final
deuteron in the $pd\to dN^*$ reaction has a form
\begin{equation}
\label{glt20}
T_{20}(\theta _{c.m.}=180^o)=
-{1\over \sqrt{2}} {w^2(q')-\sqrt{8}u(q')\,w(q')\over
u^2(q')+w^2(q')}.
\end{equation}
This formula coincides with the one for the $pd\to dp$ process within the
 NE-mechanism .

 The triangular diagram OPE  with the subprocess $pp\to d\pi^+$ was
investigated
in  \cite{12,13} in the analysis of the $pd\to dp$ process. Generalization
of the formalism from Refs. \cite{12,13} to the $pd\to dN^*$ reaction is quite
obvious if we restrict ourselves to the nucleon-like states  of $N^*$, $J^P=1/2^+$.
In this case the only difference between the
\newpage
 reactions $pd\to dN^*$ and
$pd\to dp$
is the mass inequality, $m_p\not =m_{N^*}$. Consequently, the modification of
the formalism from Refs. \cite {12,13} has a kinematic character. It results in
the following form for the spin-averaged square of the OPE amplitude
\begin{eqnarray}
\label{gl34}
{\overline {|A(pd\to dN^*)|^2}}={3\over 2}{{\widetilde G}^2\over
4\pi }\,{\widetilde F}^2(k^2)\,{E_{N^*}+m_{N^*}\over E_{N^*}^2}\,(f_{01}^2+f_{21}^2)\,
\, {3\over 2}{\overline {|A(pp\to d\pi^+)|^2}},
\end{eqnarray}
where ${\widetilde F}^2(k^2)$ is the $\pi NN^*$-formfactor; for the estimation
we use the monopole $\pi NN$-formfactor as ${\widetilde F}$; according to
Ref.\cite{14}, for the Roper resonance $N^*(1440)$ the squared coupling constant
${{\widetilde G}^2/ 4\pi }$ in the $\pi NN^*$-vertex equals $14.7\times
0.472^2$; the same value we use for the $\pi NN^*(1710)$ vertex in accordance
with arguments of Ref. \cite{3};
 $E_{N^*}$ and ${\bf p}_{N^*}$ are the total energy and momentum of the
$N^*-$isobar in the labsystem; the nuclear formfactors for the S- and D-
components of the deuteron (l=0, 2) $f_{l1}(p_{N^*})$ are expressed via
r-space  integrals of the
 product of the deuteron wave function $\psi_l(r)$ and the
spherical Bessel function of the first order, $j_1(p_{N^*}m_{N^*}\, r
/E_{N^*})$ (see details in Refs. \cite {12,13}). Such a form for $f_{l1}(
p_{N^*})$ comes from the p-wave nature  of the $\pi NN$ and $\pi NN^*(1/2^+)$
vertices. Owing to the equality $j_1(x=0)=0$, the formfactor
$f_{l1}(p_{N^*})$
becomes zero at the point $p_{N^*}=0$ and the OPE-amplitude (\ref{gl34})
vanishes, too. The rest point in the labsystem for the $N^*$-isobar is at
$T_p=1.876 GeV$ for $N^*(1440)$, 2.75 GeV for $N^*(1535)$ and 6.86 GeV
for $N^*(1710)$.

\section{Numerical results and discussion}

 The numerical calculations are performed with the Paris wave function for
the
np-component and its RGM-modification \cite{ksg} for the $NN^*$-component
of the deuteron.
The sum over ten TISM states
 listed in Tabl.2 of Ref.\cite{ksg},
 for which the effective numbers $N_{NN^*}^d$ are
not less than $10^{-5}$, is carried out in the Eq.
(\ref{gl1}) in calculation of the OBE-amplitude of $pd\to dp$ process.
The cross section of the $pd\to dN^*$ reaction is calculated  here under
 assumption that $N^*$ is a stable state to simplify the comparison with
 $pd\to dp$ process.
 The contribution of
$N^*$- exchanges to the $pd\to dp$ cross section is shown in Fig.2. The total
contribution of $N^*$-states of positive parity (s-waves) and negative
parity (p-waves) to the $pd\to dp$ cross section is by a factor of $> 30$
smaller
than the neutron exchange. In the energy interval $T_p=0.5 -1GeV$ the p-
contribution increases the OBE-cross section by a factor of $\sim 1.3$ due to
interference with the neutron exchange amplitude. However, the interference
between the s- and p-wave amplitudes of $N^*$-exchange is destructive.
As a result, the total contribution of
$N^*$-exchanges to the cross section and $T_{20}$ of the $pd\to dp$ process is
 negligible. We should note that, on the contrary, in the inclusive reaction
$d+A\to p(p^o)+X$ the interference between s- and p-waves of $N^*$-exchanges
does not occure \cite{ksg}. We found numerically  that the cross section of
 $pd\to dp$ at
$\theta _{c.m.}=180^o$, $T_p=1-3 GeV$  within
the NE-mechanism decreases by a factor $\sim 2-3$ due to rescatterings  and
 practically does not  change  its form as a function of $T_p$ (Fig.3,b). The
 tensor polarisation $T_{20}$ is   modified by the rescatterings by not more
  than  5-10\%.

 The  small contribution of $N^*-$exchanges to  $pd\to dp$ is
 mainly due to the small effective numbers of $N^*$-isobars in deuteron,
$N_{NN^*}^d <10^{-2}$. Unlike  $N^*$-exchanges in the $pd\to dp$ amplitude
including two $dNN^*$ vertices (Fig.1,a), the NE-amplitude of the $pd\to dN^*$
reaction (Fig.1,f) contains only  one $dNN^*$ vertex. Therefore the modulus of
this amplitude can be larger than that of the amplitude in Fig.1,a. Moreover,
there is an additional enhancement factor for the NE-mechanism of the
 $pd\to dN^*$
reaction in the  case of s-states of relative motion in the $d\to n+N^*$
 channel,
namely, the presence of a point with zero relative momentum ${\bf q}=0$
(\ref{gl3b}) in this channel. For the Roper resonance the point  ${\bf q}=0$
lies at $T_p=1.2 GeV$ and for $N^*(1710)$ at 2.2 GeV. It is easy to find
that the point ${\bf q}=0$ arises in the nonrelativistic kinematics, too.
For the $N^*$ isobars of negative parity the
NE-amplitude is strongly suppressed in the vicinity of the point
${\bf q}=0$ because of p-wave behaviour of the momentum distribution in the
$dnN^*$ vertex. As follows from Fig.3,{\it a}, the modulus square of
 the NE-amplitude of the
$pd\to dN^*(1710)$ reaction is the same order of magnitude as that for the
$pd\to dp$ process and  by one order of magnitude larger than the
 OPE-contribution in the energy interval of $T_p=1.5 -2 GeV$.
For the Roper resonance $N^*(1440)$ the NE-contribution is also
comparable with that for $pd\to dp$ (Fig.3,b).
 This conclusion is mainly determined
 by the effective numbers
$2\,N_{NN(1710)}^d=6.75 \,10^{-3}$ \cite{ksg}, $2\,N_{NN(1440)}^d=10^{-3}$
\cite{glozkuch} and not changed after substituting  the
harmonic oscillator wave function
 $\varphi_{00}(q) $
with the oscillator parameter
$b=0.6 fm $ \cite{15} or $b=0.8 fm$ \cite{ksg} for the RGM-modified Paris
wave function \cite {ksg}. Furthermore the NE-mechanism
of the $pd\to dN^*$ reaction can be indentified  by measurement of tensor
polarisation. We found from Eq.(\ref{glt20}) that the tensor polarisation of
the final deuteron  in the 
$pd\to dN^*$ reaction at $T_p=1-3 GeV$ is   $T_{20}\sim 0.6-0.7$ both for
the $N^*(1440)$ and $N^*(1710)$ nucleon isobar.
$T_{20}$ is approximately constant since
at energies of $T_p=1-3GeV$ the argument $q'$ in Eq.(\ref{glt20})
slowly varies in the interval of 0.7 - 0.8 GeV/c for $N^*(1440)$ and 
0.9 - 1.0 GeV/c for $N^*(1710)$.
Otherwise the tensor analyzing power of this reactions in respect  of the
initial deuteron is zero for the NE-mechanism, $t_{20}=0$.

 In accordance with the above  notes after Eq. (\ref{gl34}), the OPE mechanism
predicts a deep minimum
in the cross section of $pd\to dN^*(1440)$ at proton energy $T_p=
1.876 GeV$ (Fig.3,b) which corresponds to the rest point of $N^*(1440)$ at that
 energy. Thus, in conclusion,  there are  favourable
 conditions in the interval  $T_p=1.5 -2 GeV$
to pick out the contribution of the  NE-mechanism in the $pd\to dN^*$
reaction  for backward going  $N^*(1710)$ and $N^*(1440)$ nucleon isobars
and to search for the corresponding $NN^*$ components of the deuteron.


\section*{Acknowledgments}

The author is grateful to V.I. Komarov, A.P. Kobuchkin  and I.T. Obukhovsky
for helpful discussions. This work was supported  in part by the Russian
Foundation for
Basic Research (grants $No$ 96-02-17458 and $No$ 96-02-17215)

\eject

\eject
\baselineskip=2pc
\section*{Figure captions}
Fig.1.
The mechanisms of the $pd\to dp$  and $pd\to dN^*$
processes: the one baryon exchange (OBE)
({\it a - f}); the neutron exchange (NE) in the Born aproximation
({\it b, f}) and taking into account rescatterings ({\it c - e}); the
triangular diagram of one-pion exchange (OPE) ({\it g}).

Fig.2.
The calculated cross section of $pd\to dp$ at $\theta_{c.m.}=180^o$
as a function of kinetic energy of incident proton in the labsystem
$T_p$  within the OBE mechanisms:
  the neutron exchange (full curve, NE),
  the positive parity $N^*$
 exchange ({\it s}), the negative parity $N^*$ exchange ({\it p}),
  the total contribution of $N^*$-exchanges ({\it s}+{\it p}),
 the coherent sum  of $n+N^*$ exchanges ({\it s}+{\it p}+NE).

Fig.3.
  The  cross section of the $pd\to dN^*$ reaction at $\theta_{c.m.}=180^o$
  calculated within the different mechanisms  as a function of $T_p$
for $N^*(1710)$ ({\it a}) and  $N^*(1440)$ ({\it b}):
 curve 1 - OPE, 2 - NE.
 The  $pd\to dp$ cross section
within the NE mechanism
 is shown by curve 3 ( for the diagram in Fig 1,b)
and 4 (for the coherent sum of four diagrams in Fig.1,b - e).

%
\newpage
\begin{figure}[t]
\mbox{\epsfig{figure=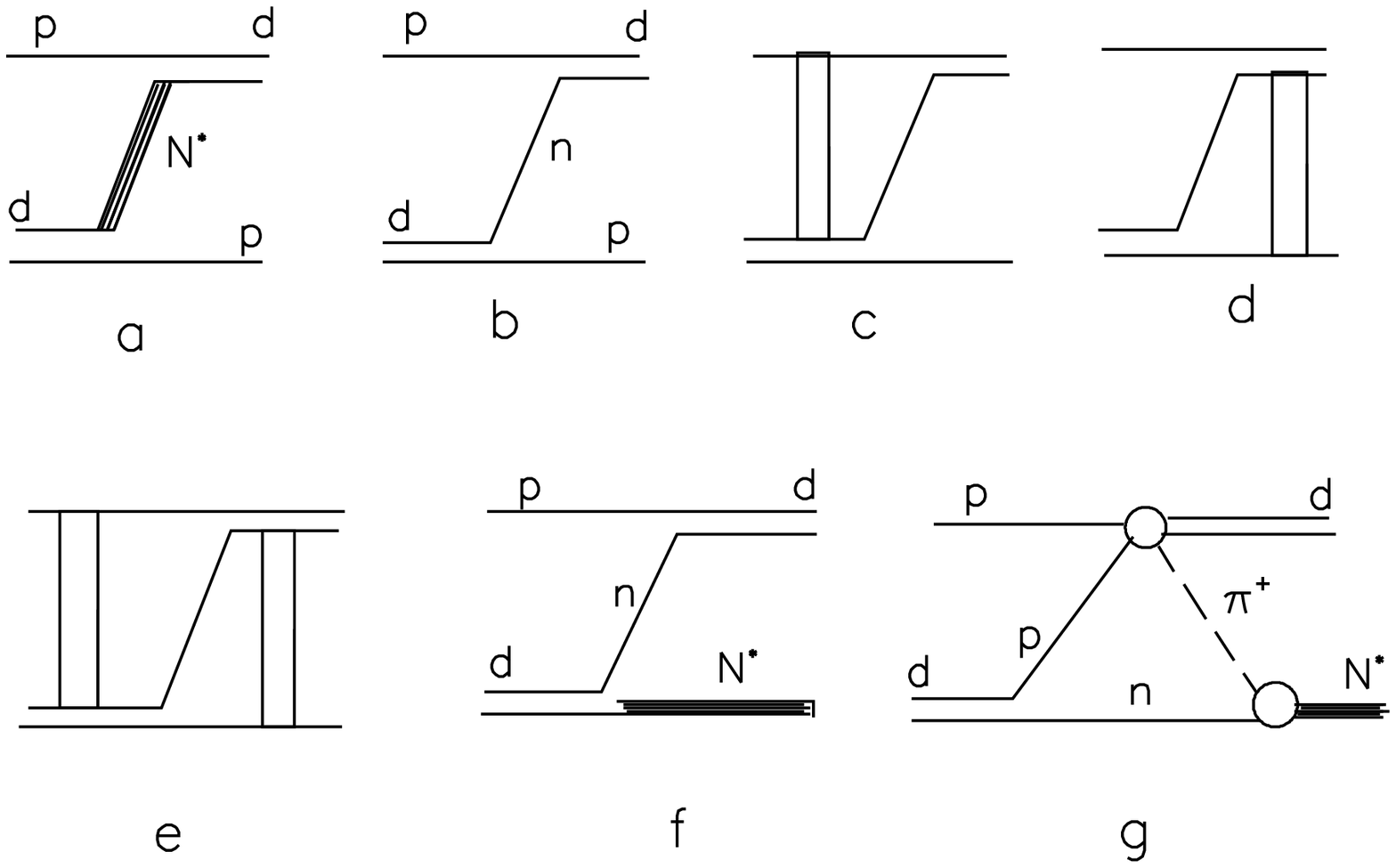,height=0.65\textheight, clip=}}
\caption{}
\end{figure}
\newpage
\begin{figure}[t]
\mbox{\epsfig{figure=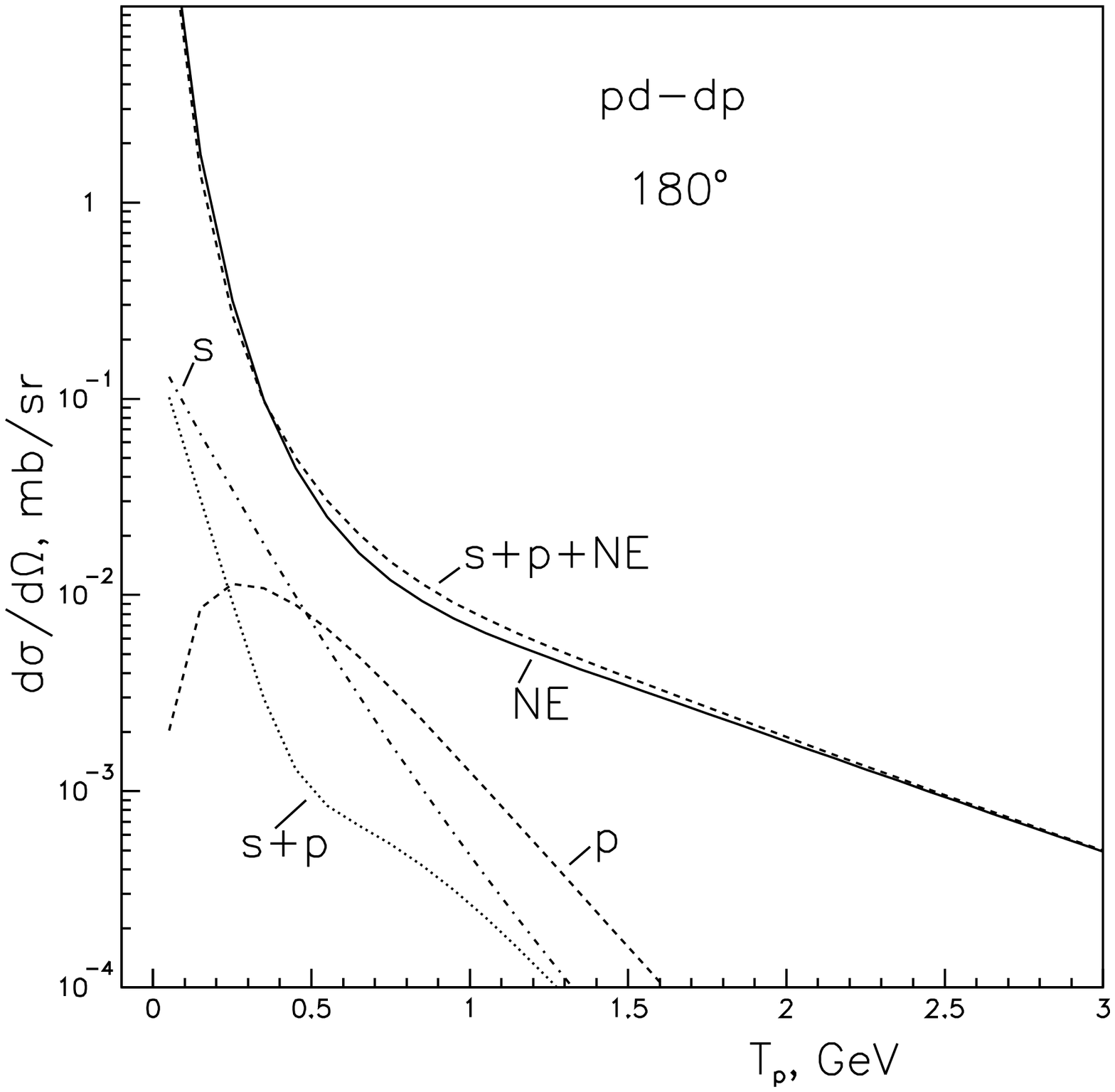,height=0.70\textheight, clip=}}
\caption{}
\end{figure}
\newpage
\begin{figure}[h]
\mbox{\epsfig{figure=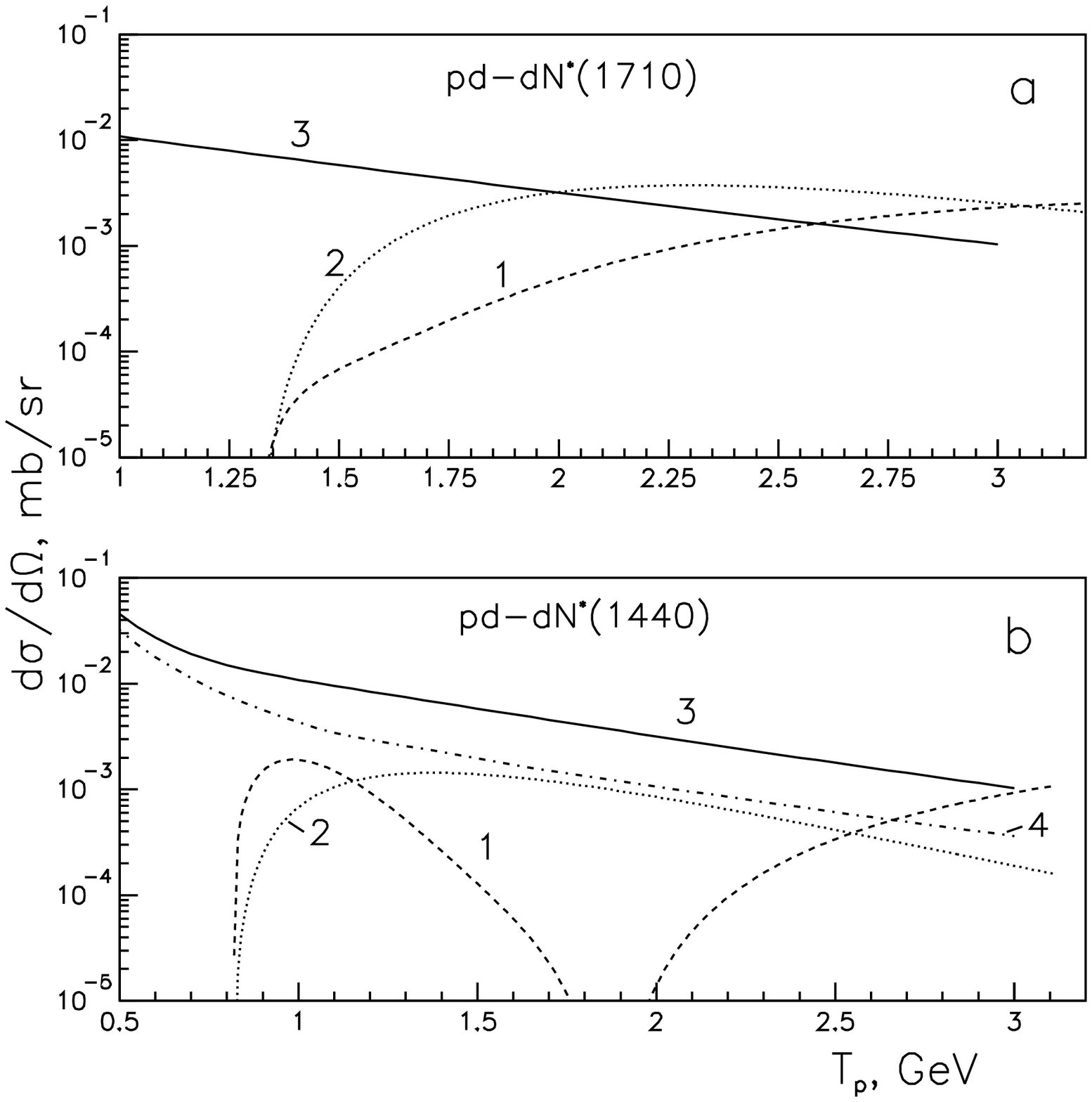,height=0.80\textheight, clip=}}
\caption{}
\end{figure}
\newpage

\end{document}